\documentclass[intlimits,twoside,a4paper]{article}
\usepackage{graphicx}
\usepackage{subcaption} 
\usepackage{float}
\usepackage{appendix}
\usepackage{listings}
\usepackage{xcolor}
\usepackage[cp1251]{inputenc}

\usepackage[eqsecnum]{cmpj3}

\DeclareMathOperator{\sech}{sech}
\DeclareMathOperator{\csch}{csch}
\DeclareMathOperator{\erfi}{erfi}

\issue{2024}{27}{4}{43301}
\doinumber{10.5488/CMP.27.43301}

\title[Eigensolutions and thermodynamic properties of generalized hyperbolic Hulthen and Woods--Saxon potential]%
{Eigensolutions and thermodynamic properties of generalized hyperbolic Hulthen and Woods--Saxon potential}%

\author[Y. M. Assimiou, S. T. Daniel, G. Issoufou,  D. F. Anselme, G. Y. H. Avossevou  ]{Y. M. Assimiou\orcid{0009-0003-0097-8607
}\refaddr{label1}\thanks{Corresponding author: \email{assimiouyaroumora@gmail.com}.}, 
S. T. Daniel\orcid{0009-0002-7603-8999}\refaddr{label1, label2},
 G. Issoufou\orcid{0000-0003-0688-6247}\refaddr{label1,label3},
 D. F. Anselme\orcid{0000-0002-2694-4144}\refaddr{label4},
 G.~Y.~H.~Avossevou\orcid{0000-0002-9609-0340}\refaddr{label1}}
 
\addresses{
\addr{label1} Institute of Mathematic and Physical Sciences (IMSP), Dangbo, Benin
\addr{label2}Ecole Polytechnique d'Abomey Calavi (EPAC-UAC), Benin
\addr{label3} D\'{e}partement de Sciences et Application, Universit\'{e} de DOSSO, Niger
\addr{label4} Facult\'{e} des Sciences et Techniques, Universit\'{e} Nationale des Sciences, Technologies,\\ Ing\'{e}nierie et Math\'{e}matiques (UNSTIM), Abomey, Benin
}


\date{Received June 29, 2024, in final form October 08, 2024}

\begin{document}
\maketitle
\begin{abstract}
In this paper, we present the solutions of the Schr\"{o}dinger equation and the thermodynamic properties of generalized hyperbolic Hulthen and Woods--Saxon potential.  The eigenvalues and eigenfunctions were found using the parametric Nikiforov--Uvarov  method (PNUM). The clean energies of the molecules HCl, NiC, CO, I$_2$, H$_2$, LiH, CuLi and CrH are calculated for certain values of $n$ and $\ell$. They are positive and close to the energy of the ground state ($n= \ell= 0$) in the case of the atomic unit (whose energies become negative for $n=2$). The figures show that the proper energies decrease as $n$, $\ell$, $\alpha$ increase, while they increase as $m$ increases, which confirms the results obtained in the literature. The obtained energy  was used to calculate the partition function from
which thermodynamic properties such as average energy, specific heat capacity, entropy
and free energy are calculated. Numerical results are generated for this generalized hyperbolic Hulthen and Woods--Saxon  potential. This study showed that the disorder decreases if the temperature decreases and this decrease is more rapid for HCl and H$_{2}$ molecules.

\printkeywords
%
\end{abstract}

\section{Introduction}
To be aware of all the possible information of a physical system, and to ensure its appropriate
description, we must determine in the form of a wave function an analytical solution having all the important properties~\cite{D1}.
In determining this analytical solution in the form of a wave function, the Schr\"{o}dinger equation plays a primordial role,
very fundamental and capital because it shows and governs the evolvement of the physical system over time. Several physicist researchers have sought to find analytical solutions of
the Schr\"{o}dinger equation for several physical systems after Schr\"{o}dinger's work~\cite{D7}.

The eigenenergy obtained is used to calculate the partition function which is the central parameter for studying the thermodynamic properties: the mean energy, specific heat capacity (thermal capacity), entropy and free energy.  These thermodynamic properties constitute one of the essential elements of quantum physics. Several potential models have been widely reported lately since recently many researchers have got interested in the study of thermodynamic properties given their applications and properties \cite{As7}. For example, Dong and Cruz-Irisson
calculated the thermodynamic properties of the modified Rosen--Morse potential~\cite{As10}. Khordad and Sedehi \cite{As12}, studied the different thermodynamic properties for double ring-shape potential. Inyang et al. \cite{As13}, calculated the various thermodynamic properties of Eckart--Hellman potential in one of the recent studies. 
 Njoku et al. \cite{As16} in their study obtained approximate solutions of Schr\"{o}dinger equation
and thermodynamic properties with Hua potential. Demirci and 
Sever \cite{D51},  calculated the nonrelativistic thermal properties of Eckart plus 
class of Yukawa potential. Ramantswana et al. \cite{As19}, in their own study, determined the 
thermodynamic properties of CrH, NiC and CuLi diatomic molecules with the linear 
combination of Hulthen-type potential plus Yukawa potential. The thermodynamic properties of 
some diatomic molecules confined within anharmonic oscillating system were also studied 
by Oluwadare et al. \cite{As20}. Wang et al. \cite{As21}, in a specific form, predicted the ideal-gas thermodynamic 
properties for water and deduced the average relative deviations of the predicted reduced molar 
Gibbs free energy and molar entropy of water from NIST data.
In reference~\cite{A27}, Okon et al. studied thermodynamic properties using hyperbolic Hulthen plus hyperbolic 
exponential inversely quadratic potential. Edet et al. \cite{A28}, used Poisson summation 
approach to study thermal properties of Deng-fan Eckart potential model. Ikot et al.~\cite{A29}, investigated the 
thermodynamic properties of diatomic molecules with general molecular potential. 
 Okorie et al. \cite{A31}, calculated the thermodynamic properties of the modified Yukawa 
potential.
Okorie et al. \cite{A32}, studied thermodynamic properties with modified Mobius square potential (MMSP). 
Okon et al. \cite{A23}, combined Mobius Square and Screened Kratzer Potential to obtain 
thermodynamic properties and bound State Solutions to Schr\"{o}dinger equation.
Omugbe et al. \cite{A34},  studied thermodynamic properties 
and expectation values of a mixed hyperbolic Poschl-Teller potential (MHPTP). Isonguyo~\cite{D81} studied eigensolutions and thermodynamic properties of Kratzer plus generalized Morse potential. Okon et al. \cite{A}, obtained eigensolution and thermodynamic properties of Manning Rosen plus exponential Yukawa 
potential. Emeje et al. \cite{As}, examined eigensolution and thermodynamic properties of
standard Coulombic potential, etc.

A linear potential combination can be used to examine the interactions between unshaped pairs of the nucleus and the spin-orbit coupling inside the potential fields, as well as to determine the vibrations of hadronic systems and create a suitable model for other physical phenomena \cite{D51}.
Generalized hyperbolic potentials of Hulthen and Woods--Saxon  are used in several major fields such as: nuclear physics, atomic, molecular, quantum, statistical, condensed matter and quantum chemistry. These potentials are useful mathematical tools for understanding and predicting the behavior of quantum systems in a variety of contexts, from nuclear physics to quantum chemistry \cite{D51}.

However, to our knowledge, generalized hyperbolic Hulthen and Woods--Saxon potential is not yet studied. Based on the above motivations, being
interested in thermodynamic properties, taking into consideration  previous studies and in order to enrich previous attempts, we propose a new combined potential obtained by a linear combination of the generalized hyperbolic, Hulthen and Woods--Saxon potential [$V_{\text{GHHWSP}}(r)$].
This potential can be favourably used in several theoretical and practical areas. In nuclear physics~\cite{E77}: for  modelling the nuclear structure, for the analysis of nuclear atomic properties, and for the study of interactions between nucleons. In atomic physics \cite{E77}: for  understanding the interactions between electrons and the nuclei, modelling the energy levels of atoms. In statistical physics  \cite{E77}: for  modeling the particles systems, and for the study of thermodynamic and phase properties. In condensed matter physics  \cite{E80}: for  modelling the electronic structure, for the study of the band properties, for simulation of transport properties, and to describe a system of interactions including bound and continuous states, as well as the electromagnetic transition. In molecular physics  \cite{As}: for the prediction of the spectroscopic properties of molecules, for the analysis of interactions between atoms in molecules. In quantum physics  \cite{D51}:
for the study of particle diffusion in various potentials, for modelling the bound states of quantum systems, for the analysis of spectral properties of quantum systems. In quantum chemistry  \cite{D51, As19, A28}: for the calculation of energies and electronic structures of molecules, and to predict the chemical reactions and compounds properties.

Our goal is to examine it later on in several theoretical and practical areas. For now, we determine the solutions to the bound state of the Schr\"{o}dinger equation of this potential using the parametric Nikiforov--Uvarov method (PNUM) and the Greene-Aldrich approximation scheme.  
We present the eigenvalues of the bound states and the corresponding wave eigenfunctions. We also discuss some special cases and the thermodynamic properties of the potential $V_{\text{GHHWSP}}(r)$.

Note that the Schr\"{o}dinger equation is only possible with the kinetic moment $\ell=0$ for some potential  models. However, for the states $\ell\neq0$, it is necessary to use some approximations such as the Pekeris approximation, in order to treat the centrifugal orbit term or solve it numerically \cite{D53}.

The article is organized as follows: section~\ref{sec:2} is devoted to the description of the parametric Nikiforov--Uvarov method (PNUM). In section~\ref{sec:3}, we solve the Schr\"{o}dinger equation with the potential $V_{\text{GHHWSP}}(r)$. In section~\ref{sec:4} some special cases are presented. Thermodynamic properties are presented in section~\ref{sec:5}. The numerical results and discussions are presented in section~\ref{sec:6},  followed by the conclusion in section~\ref{sec:concl}.

\section{Parametric Nikiforov--Uvarov method (PNUM)}
\label{sec:2}
The Nikiforov--Uvarov method (NUM) is a powerful method for reducing the differential equations of the order 2 with an appropriate transformation  $x=x(s)$  to the hypergeometric type of the following form:
\begin{eqnarray}
\Psi''(s) +\frac{\tilde{\tau}(s)}{\sigma(s)}\Psi'(s)+\frac{\tilde{\sigma}(s)}{\sigma^{2}(s)}\Psi(s)=0,\label{1}
\end{eqnarray} where $ \sigma(s) $ and $ \tilde{\sigma}(s) $ are polynomials of degree at most 2, while $ \tilde{\tau}(s) $ is a polynomial of degree at most~1~\cite{D55}.

The generalization of the parametric Nikiforov--Uvarov method (PNUM) is given by the general differential equation \cite{D81}: 
\begin{eqnarray}
\Psi''(s) +\frac{c_{1}-c_{2}s}{s(1-c_{3}s)}\Psi'(s)+\frac{[-\varepsilon_{1}s^{2}+\varepsilon_{2}s-\varepsilon_{3}]}{s^{2}(1-c_{3}s)^{2}}\Psi(s)=0,\label{14}
\end{eqnarray}
for its resolution, it is compared to the equation (\ref{1}) and we have: 
\begin{eqnarray}
\tilde{\tau}(s)=c_{1}-c_{2}s,\quad \sigma(s)=s(1-c_{3}s),\quad \tilde{\sigma}(s)=-\varepsilon_{1}s^{2}+\varepsilon_{2}s-\varepsilon_{3}. \label{15}
\end{eqnarray}
The determination of the parameters of equation (\ref{14}) plays a very important role as tools to find the eigenvalues and eigenfunctions of the  energy.

The other parameters are:
\begin{align}
c_{4}&=\frac{1}{2}(1-c_{1}),\quad c_{5}=\frac{1}{2}(c_{2}-2c_{3}),\quad c_{6}=c_{5}^{2}+\varepsilon_{1},\nonumber\\ c_{7}&=2c_{4}c_{5}-\varepsilon_{2},\quad c_{8}=c_{4}^{2}+\varepsilon_{3},\quad c_{9}=c_{3}c_{7}+c_{3}^{2}c_{8}+c_{6},\label{17}\nonumber\\
c_{10}&=c_{1}+2c_{4}+2\sqrt{c_{8}},\quad c_{11}=c_{2}-2c_{5}+2(c_{3}\sqrt{c_{8}}+\sqrt{c_{9}}),\nonumber\\
c_{12}&=c_{4}+\sqrt{c_{8}}, \quad c_{13}=c_{5}-(c_{3}\sqrt{c_{8}}+\sqrt{c_{9}}).
\end{align}
The equation of the energy spectrum is given by:
\begin{eqnarray}
c_{2}n-(2n+1)c_{5}+n(n-1)c_{3}+(2n+1)(c_{3}\sqrt{c_{8}}+\sqrt{c_{9}})+c_{7}+2c_{3}c_{8}+2\sqrt{c_{8}c_{9}}=0.\label{22}
\end{eqnarray}
The total wave function is therefore given by:
\begin{eqnarray}
\Psi_{n\ell}(s)=C_{n\ell}s^{c_{12}}(1-c_{3}s)^{-c_{12}-({c_{13}}/{c_{3}})}P_{n}^{(c_{10}-1,\,({c_{11}}/{c3})-c_{10}-1)}(1-2c_{3}s),\label{28}
\end{eqnarray} 
with $C_{n\ell}$ being the normalization constant and $P_{n}^{(\alpha,\beta)}(1-2c_{3}s)$ are the Jacobi polynomials.

\section{Bound state solution of the Schr\"{o}dinger equation}
\label{sec:3}
In spherical coordinates, the radial part of the Schr\"{o}dinger wave is in the form  \cite{D41}:
\begin{eqnarray}
\frac{\rd^{2}\Psi_{n\ell}(r)}{\rd r^{2}}+
\frac{2m}{\hbar^{2}}\left[E_{n\ell}-V(r)-\frac{\hbar^{2}\ell(\ell+1)}{2mr^{2}}\right]\Psi_{n\ell}(r)=0, \label{a2}
\end{eqnarray}where $\Psi_{n\ell}(r)$ is the radial wave function, $E_{n\ell}$ is the energy spectrum, $\hbar$ is 
the reduced Planck’s constant, $m$ is the reduced mass, ${\ell(\ell+1)}/{r^{2}}$ is the 
centrifugal barrier, which is resolved by the Greene-Aldrich approximation scheme, that is a good approximation for $\alpha r \ll 1$, given by  \cite{D69}:
\begin{eqnarray}
\frac{1}{r^{2}}\approx 4\alpha^{2}\frac{\re^{-2\alpha r}}{(1-\re^{-2\alpha r})^{2}},
\label{a5}
\end{eqnarray}
 $\ell$ and $n$ being the angular momentum and 
principal quantum numbers, respectively, and $V(r)$ represents here the generalized hyperbolic Hulthen and Woods--Saxon potential denoted by $V_{\text{GHHWSP}}(r)$ given by:
\begin{align}
V_{\text{GHHWSP}}(r) = &V_{1}\tanh^{2}(\alpha r)+V_{2}\coth^{2}(\alpha r)-V_{3}\sech^{2}(\alpha r)-V_{4}\csch^{2}(\alpha r)\nonumber\\
&-\frac{V_{0}\re^{-2\alpha r}}{(1-\re^{-2\alpha r})^{2}}-\frac{V_{5}\re^{-2\alpha r}}{(1+\re^{-2\alpha r})^{2}},\label{a1}
\end{align}
where $V_{0}$, $V_{1}$, $V_{2}$, $V_{3}$, $V_{4}$ and $V_{5}$, are potential parameters having the dimension of an energy, also $V_{0}$ and $V_{5}$ representing the depths of potential wells and $\alpha$ is the range of the potential.

Using equation (\ref{a5}), the approximate potential of equation (\ref{a1}), is given by:
\begin{eqnarray}
V'_{\text{GHHWSP}}(r)=V_{1}\tanh^{2}(\alpha r)+V_{2}\coth^{2}(\alpha r)-\frac{V_{5}+4V_{3}}{4}\sech^{2}(\alpha r)-\frac{V_{0}+4V_{4}}{4}\csch^{2}(\alpha r).\label{aa1}
\end{eqnarray}
By substituting the equations (\ref{a5}) and (\ref{aa1})  
into (\ref{a2}) we have:
\begin{align}
&\frac{\rd^{2}\Psi_{n\ell}(r)}{\rd r^{2}}+
\frac{2m}{\hbar^{2}}\bigg[E_{n\ell}-V_{1}\tanh^{2}(\alpha r)-V_{2}\coth^{2}(\alpha r)\label{a6}\\
&+\alpha_{1}\sech^{2}(\alpha r)+\alpha_{2}\coth^{2}(\alpha r)\bigg]\Psi_{n\ell}(r)=0,\nonumber 
\end{align}
with $\alpha_{1}=({V_{5}+4V_{3}})/{4}$ and $\alpha_{2}=({V_{0}+4V_{4}})/{4}-[{\hbar^{2}\ell(\ell+1)\alpha^{2}}/{2m}]$.

By introducing the substitution $s=\tanh^{2}(\alpha r)$, we have obtained the following second order differential equation:
\begin{eqnarray}
\Psi''(s)+\frac{\big(\frac{1}{2}-\frac{3}{2}s\big)}{s(1-s)}\Psi'(s)+\frac{1}{\left[s(1-s)\right]^{2}}\left[-\varepsilon_{1}s^{2}+\varepsilon_{2}s-\varepsilon_{3}\right]\Psi(s)=0,\label{81}
\end{eqnarray}
with the new parameters:
\begin{align} \label{aa2}
	\varepsilon&=\frac{2m}{\hslash^{2}\alpha^{2}}E_{n\ell}, \quad
	\varepsilon_{1}=\frac{2m}{4\hslash^{2}\alpha^{2}}(V_{1}+\alpha_{1}), \quad \varepsilon_{2}=\frac{2m}{4\hslash^{2}\alpha^{2}}(E_{n\ell}+\alpha_{1}-\alpha_{2}), \nonumber\\ \varepsilon_{3}&=\frac{2m}{4\hslash^{2}\alpha^{2}}(V_{2}-\alpha_{2}).
\end{align} 
We note that equation (\ref{81}) has a suitable form for implementing the parametric Nififorov-Uvarov method.
From the equations (\ref{1}), (\ref{14}), (\ref{15}) and (\ref{81}), we have by identification:
\begin{eqnarray}
\tilde{\tau}(s)=c_{1}-c_{2}s=\frac{1}{2}-\frac{3}{2}s, \quad \sigma(s)=s(1-c_{3}s)=s(1-s)\label{83},\quad \tilde{\sigma}(s)=-\varepsilon_{1}s^{2}+\varepsilon_{2}s-\varepsilon_{3}.
\end{eqnarray}
And then according to the equations (\ref{17}) and (\ref{83}), we have obtained a parameter set:
\begin{align}
c_{1}&=\frac{1}{2}, \quad c_{2}=\frac{3}{2}, \quad c_{3}=1, \quad c_{4}=\frac{1}{4}, \quad c_{5}=-\frac{1}{4}, \quad c_{6}=\frac{1}{16}+\varepsilon_{1},\nonumber\\ 
c_{7}&=-\frac{1}{8}-\varepsilon_{2}, \quad c_{8}=\frac{1}{16}+\varepsilon_{3}, \quad c_{9}=\varepsilon_{1}-\varepsilon_{2}+\varepsilon_{3}, \quad c_{10}=1+2\sqrt{\frac{1}{16}+\varepsilon_{3}},\label{84}\nonumber\\ c_{11}&=2+2\left(\sqrt{\varepsilon_{1}-\varepsilon_{2}+\varepsilon_{3}}+\sqrt{\frac{1}{16}+\varepsilon_{3}}\right), \quad c_{12}=\frac{1}{4}+\sqrt{\frac{1}{16}+\varepsilon_{3}}, \nonumber\\ c_{13}&=-\frac{1}{4}-\left(\sqrt{\varepsilon_{1}-\varepsilon_{2}+\varepsilon_{3}}+\sqrt{\frac{1}{16}+\varepsilon_{3}}\right). 
\end{align}
According to equations (\ref{22}) and (\ref{84}), the energy equation is obtained from:
\begin{align}
&\left(n^{2}+n+\frac{1}{4}\right)+(2n+1)\left(\sqrt{\varepsilon_{1}-\varepsilon_{2}+\varepsilon_{3}}+\sqrt{\frac{1}{16}+\varepsilon_{3}}\right)+(2\varepsilon_{3}-\varepsilon_{2})\nonumber\\
&+2\sqrt{\left(\frac{1}{16}+\varepsilon_{3}\right)(\varepsilon_{1}-\varepsilon_{2}+\varepsilon_{3})}=0.\label{91}
\end{align}
Finally, according to equations (\ref{aa2}) and (\ref{91}), we obtained
the following expression of the energy eigenvalue:
\begin{align}
E_{n\ell}= &(V_{1}+V_{2})-\frac{\hslash^{2}\alpha^{2}}{2m}\bigg[(2n+1)+\sqrt{\frac{1}{4}+\frac{2m}{4\hslash^{2}\alpha^{2}}(4V_{2}-4V_{4}-V_{0})+\ell(\ell+1)}\nonumber\\
&\mp\sqrt{\frac{1}{4}+\frac{2m}{4\hslash^{2}\alpha^{2}}(4V_{1}+4V_{3}+V_{5})}\bigg]^{2},\label{94}
\end{align}
with
${1}/{4}+\left({2m}/{4\hslash^{2}\alpha^{2}}\right)(4V_{2}-4V_{4}-V_{0})+\ell(\ell+1)\geqslant 0$.

According to equations (\ref{28}), and (\ref{84}), the total wave function is given as:
\begin{eqnarray}
\Psi_{n\ell}(s)=C_{n\ell}s^{{1}/{4}+\sqrt{({1}/{16})+\varepsilon_{3}}}(1-s)^{\sqrt{\varepsilon_{1}-\varepsilon_{2}+\varepsilon_{3}}}P_{n}^{2\sqrt{({1}/{16})+\varepsilon_{3}}, \; 2\sqrt{\varepsilon_{1}-\varepsilon_{2}+\varepsilon_{3}}}(1-2s),\label{98}
\end{eqnarray} 
with $s=\tanh^{2}(\alpha r)$,  $C_{n\ell}$ the normalization constant and $P_{n}^{2\sqrt{({1}/{16})+\varepsilon_{3}}, \; 2\sqrt{\varepsilon_{1}-\varepsilon_{2}+\varepsilon_{3}}}$ are the Jacobi polynomials.

\section{Some special cases}
\label{sec:4}

\begin{enumerate}
\item If $V_{0}=V_{5}=0$, we find ourselves at the generalized hyperbolic potential proposed by Okorie et al., and given by the equation (1) of \cite{D41}:
\begin{eqnarray}
V_{\text{GHP}}(r)=V_{1}\tanh^{2}(\alpha r)+V_{2}\coth^{2}(\alpha r)-V_{3}\sech^{2}(\alpha r)-V_{4}\csch^{2}(\alpha r),\label{99}
\end{eqnarray} the expression of the equation (\ref{94}) in this case becomes:
\begin{align}\label{101}
&E_{n\ell}=(V_{1}+V_{2})\nonumber\\
&-\frac{\hslash^{2}\alpha^{2}}{2m}\left\{ 2n+\frac{1}{2}+\frac{1}{2}\left[1+\sqrt{1+\frac{8m}{\hslash^{2}\alpha^{2}}(V_{2}-V_{4})+(2\ell+1)^{2}}\right]\mp\sqrt{\frac{1}{4}+\frac{2m}{\hslash^{2}\alpha^{2}}(V_{1}+V_{3})}\right\}^{2},
\end{align}
which conforms to equation (25) of \cite{D41}.

\item If $V_{0}=V_{2}=V_{4}=V_{5}=0$, we find ourselves at the hyperbolic Rosen Morse potential given by equation (26) of \cite{D41}:
\begin{eqnarray}
V_{\text{RMHP}}(r)=V_{1}\tanh^{2}(\alpha r)-V_{3}\sech^{2}(\alpha r),\label{102}
\end{eqnarray} the expression of the equation (\ref{94}) in this case becomes:
\begin{eqnarray}
E_{n\ell}=V_{1}-\frac{\hslash^{2}\alpha^{2}}{2m}\left[\left(2n+\ell+\frac{3}{2}\right)\mp\sqrt{\frac{1}{4}+\frac{2m}{\hslash^{2}\alpha^{2}}(V_{1}+V_{3})}\right]^{2},\label{103}
\end{eqnarray}
which conforms to equation (27) of \cite{D41}.

\item If $V_{0}=V_{1}=V_{3}=V_{5}=0$, we find ourselves at the hyperbolic Eckart potential with the form given by equation (28) of \cite{D41}:
\begin{eqnarray}
V_{\text{EHP}}(r)=V_{2}\coth^{2}(\alpha r)-V_{4}\csch^{2}(\alpha r),\label{104}
\end{eqnarray} the expression of the equation (\ref{94}) in this case becomes:
\begin{eqnarray}
E_{n\ell}=V_{2}-\frac{\hslash^{2}\alpha^{2}}{2m}\left\{(2n+1)+\frac{1}{2}\left[\sqrt{1+\frac{8m}{\hslash^{2}\alpha^{2}}(V_{2}-V_{4})+(2\ell+1)^{2}}\mp1\right]\right\}^{2},\label{105}
\end{eqnarray}
which conforms to  equation (27) of \cite{D41}.

\item If $V_{5}=0$, we find ourselves at a combined potential by the generalized hyperbolic potential plus Hulthen of the form:
\begin{eqnarray}
V_{\text{GHHP}}(r)=V_{1}\tanh^{2}(\alpha r)+V_{2}\coth^{2}(\alpha r)-V_{3}\sech^{2}(\alpha r)-V_{4}\csch^{2}(\alpha r)-\frac{V_{0}\re^{-2\alpha r}}{(1-\re^{-2\alpha r})^{2}},\label{106}
\end{eqnarray} the expression of the equation (\ref{94}) in this case becomes:
\begin{align}\label{107}
&E_{n\ell}=(V_{1}+V_{2})\nonumber\\
&-\frac{\hslash^{2}\alpha^{2}}{2m}\left[(2n+1)+\sqrt{\frac{1}{4}+\frac{2m}{4\hslash^{2}\alpha^{2}}(4V_{2}-4V_{4}-V_{0})+\ell(\ell+1)}\mp\sqrt{\frac{1}{4}+\frac{2m}{\hslash^{2}\alpha^{2}}(V_{1}+V_{3})}\right]^{2}.
\end{align}

\item If $V_{0}=0$, we find ourselves at a combined potential by the generalized hyperbolic potential plus Woods--Saxon potential of the form:
\begin{eqnarray}
V_{\text{GHWSP}}(r)=V_{1}\tanh^{2}(\alpha r)+V_{2}\coth^{2}(\alpha r)-V_{3}\sech^{2}(\alpha r)-V_{4}\csch^{2}(\alpha r)-\frac{V_{5}\re^{-2\alpha r}}{(1+\re^{-2\alpha r})^{2}},\label{108}
\end{eqnarray} and the expression of the equation (\ref{94}) in this case becomes:
\begin{align}\label{109}
E_{n\ell}={}&(V_{1}+V_{2})
-\frac{\hslash^{2}\alpha^{2}}{2m}\bigg[2n+1\nonumber\\
{}&+\sqrt{\frac{1}{4}+\frac{2m}{\hslash^{2}\alpha^{2}}(V_{2}-V_{4})+\ell(\ell+1)}\mp\sqrt{\frac{1}{4}+\frac{2m}{4\hslash^{2}\alpha^{2}}(4V_{1}+4V_{3}+V_{5})}\bigg]^{2}.
\end{align}

\item If $V_{1}=V_{2}=V_{3}=V_{4}=0$, we find ourselves at a combined potential by the Hulthen plus Woods--Saxon one of the form:
\begin{eqnarray}
V_{\text{HWSP}}(r)=-\frac{V_{0}\re^{-2\alpha r}}{(1-\re^{-2\alpha r})^{2}}-\frac{V_{5}\re^{-2\alpha r}}{(1+\re^{-2\alpha r})^{2}},\label{110}
\end{eqnarray} and  the expression of the equation (\ref{94}) in this case becomes:
\begin{eqnarray}
E_{n\ell}=-\frac{\hslash^{2}\alpha^{2}}{2m}\left[(2n+1)+\sqrt{\frac{1}{4}-\frac{2m}{4\hslash^{2}\alpha^{2}}V_{0}+\ell(\ell+1)}\mp\sqrt{\frac{1}{4}+\frac{2m}{4\hslash^{2}\alpha^{2}}V_{5}}\right]^{2}.\label{111}
\end{eqnarray}

\item If $V_{0}=V_{3}=V_{4}=V_{5}=0$, we find ourselves at a hyperbolic potential of the form:
\begin{eqnarray}
V_{\text{HP}}(r)=V_{1}\tanh^{2}(\alpha r)+V_{2}\coth^{2}(\alpha r),\label{112}
\end{eqnarray} and the expression of the equation (\ref{94}) in this case becomes:
\begin{eqnarray}
E_{n\ell}=(V_{1}+V_{2})-\frac{\hslash^{2}\alpha^{2}}{2m}\left[(2n+1)+\sqrt{\frac{1}{4}+\frac{2m}{\hslash^{2}\alpha^{2}}V_{2}+\ell(\ell+1)}\mp\sqrt{\frac{1}{4}+\frac{2m}{\hslash^{2}\alpha^{2}}V_{1}}\right]^{2}.\label{113}
\end{eqnarray}

\item If $V_{3}=V_{4}=0$, we find ourselves at a  given potential by the linear combination of the hyperbolic Hulthen and Woods--Saxon potential of the form:
\begin{eqnarray}
V_{\text{HHWSP}}(r)=V_{1}\tanh^{2}(\alpha r)+V_{2}\coth^{2}(\alpha r)-\frac{V_{0}\re^{-2\alpha r}}{(1-\re^{-2\alpha r})^{2}}-\frac{V_{5}\re^{-2\alpha r}}{(1+\re^{-2\alpha r})^{2}},\label{114}
\end{eqnarray}
and the expression of the equation (\ref{94}) in this case becomes:
\begin{align}\label{115}
&E_{n\ell}=V_{1}+V_{2}\nonumber\\
&-\frac{\hslash^{2}\alpha^{2}}{2m}\left[2n+1+\sqrt{\frac{1}{4}+\frac{2m}{4\hslash^{2}\alpha^{2}}(4V_{2}-V_{0})+\ell(\ell+1)}\mp\sqrt{\frac{1}{4}+\frac{2m}{4\hslash^{2}\alpha^{2}}(4V_{1}+V_{5})}\right]^{2}.
\end{align}

\item If $V_{1}=V_{2}=0$ we find ourselves at a  given potential by the linear combination of the hyperbolic Hulthen and Woods--Saxon potential of the form:
\begin{eqnarray}
V_{\text{HHWSP}}(r)=-V_{3}\sech^{2}(\alpha r)-V_{4}\csch^{2}(\alpha r)
-\frac{V_{0}\re^{-2\alpha r}}{(1-\re^{-2\alpha r})^{2}}-\frac{V_{5}\re^{-2\alpha r}}{(1+\re^{-2\alpha r})^{2}},\label{116}
\end{eqnarray}
and the expression of the equation (\ref{94}) in this case becomes:
\begin{eqnarray}
E_{n\ell}=
-\frac{\hslash^{2}\alpha^{2}}{2m}\left[2n+1+\sqrt{\frac{1}{4}-\frac{2m}{4\hslash^{2}\alpha^{2}}(4V_{4}+V_{0})+\ell(\ell+1)}\mp\sqrt{\frac{1}{4}+\frac{2m}{4\hslash^{2}\alpha^{2}}(4V_{3}+V_{5})}\right]^{2}.\label{117}
\end{eqnarray}
	 
\end{enumerate}

\section{Non-relativistic thermodynamic properties}
\label{sec:5}

Several thermodynamic properties can be studied from the partition function defined as:
\begin{eqnarray}
Z(\beta)=\sum_{n=0}^{\lambda}\re^{-\beta E_{n\ell}} \label{e1},
\end{eqnarray} 
where $ \lambda$ is the largest value of the vibrational quantum number obtained from the numerical solution \cite{D81}: $\frac{\rd E_{n\ell}}{\rd n}=0$, $\beta=\frac{1}{kT}$, where $k$ and $T$ are the Botzmann constants and the absolute temperature, respectively. The summation of (\ref{e1}) can be replaced by the integral in the classical limit:
\begin{eqnarray}
Z(\beta)=\int_{0}^{\lambda}\re^{-\beta E_{n\ell}}\, \rd n.\label{T59}
\end{eqnarray} 
The expression of the equation (\ref{94}) can be written in the form:
\begin{eqnarray}
E_{n\ell}=V-q[2n+Q]^{2},\label{T60}
\end{eqnarray} 
where 
\begin{align}\label{T61}
V&=(V_{1}+V_{2}), \quad q=\frac{\hslash^{2}\alpha^{2}}{2m},\nonumber\\
Q&=\sqrt{\frac{1}{4}+\frac{2m}{4\hslash^{2}\alpha^{2}}(4V_{2}-4V_{4}-V_{0})+\ell(\ell+1)}\mp\sqrt{\frac{1}{4}+\frac{2m}{4\hslash^{2}\alpha^{2}}(4V_{1}+4V_{3}+V_{5})}.
\end{align}
The partition function equation (\ref{T59}) can be expressed in the classical limit as:
\begin{eqnarray}\label{T62}
Z(\beta)=\int_{0}^{\lambda}\re^{-\beta\left(V-q[2n+Q]^{2}\right)}\, \rd n.
\end{eqnarray}
The equation (\ref{T62}) was integrated with the help of Mathematica and the partition function becomes:
\begin{eqnarray}\label{T63}
Z(\beta)=\frac{\re^{-V\beta}\sqrt{\piup}\left(-\erfi[Q\sqrt{q\beta}]+\erfi[(Q+2\lambda)\sqrt{q\beta}]\right)}{4\sqrt{q\beta}},
\end{eqnarray}
with $q>0$, $\lambda>0$ and $\erfi(z)=-\ri\erfi(\ri z)=\sqrt{\frac{4}{\piup}}\int_{0}^{z}\re^{u^{2}}\rd u$~\cite{D115}. 

\begin{enumerate}
\item  \textbf{Average energy}\\
Mean energy is given by:
\begin{align}
U(\beta)=&-\frac{\partial \ln Z(\beta)}{\partial\beta},\nonumber\\
U(\beta)={}&\ln\big\{-2\sqrt{q\beta}\big[\re^{Q^{2}q\beta}Q-\re^{q\beta(Q+2\lambda)^{2}}(Q+2\lambda)\big]\nonumber\\
{}&+\sqrt{\piup}(1+2V \beta)\erfi[Q\sqrt{q\beta}]-\sqrt{\piup}(1+2V \beta)\erfi[(Q+2\lambda)\sqrt{q\beta}]\big\}/2\beta\lambda_{1},
\end{align}
with $\lambda_{1}=\sqrt{\piup}\left\{\erfi[Q\sqrt{q\beta}]-\erfi[(Q+2\lambda)\sqrt{q\beta}]\right\}$.

\item \textbf{Specific heat capacity}\\
Specific heat capacity is given by:
\begin{align}
C(\beta)={}&k_{\text{B}}\beta^{2}\frac{\partial^{2} \ln Z(\beta)}{\partial\beta^{2}}=-k_{\text{B}}\beta^{2}\frac{\partial \ln U(\beta)}{\partial\beta},\nonumber\\
C(\beta)={}&k_{\text{B}}\Big\{-2 \re^{2 q Q^2 \beta}
q \beta \left[(-1 + \re^{4 q \beta \lambda (Q + \lambda)}) Q + 
2\lambda \re^{4 q \beta \lambda (Q + \lambda)} \right]^2 + \
\piup \erfi[ Q \sqrt{q\beta}]^2\nonumber\\
{}&-\sqrt{\piup} \sqrt{q\beta} \left[\re^{q Q^2 \beta}
Q (-1 + 2 q Q^2 \beta) - 
\re^{q \beta (Q + 2 \lambda)^2} (Q + 2 \lambda) \Big(-1 + 
2 q \beta (Q + 2 \lambda)^2\Big)\right]\nonumber\\
{}&\times \erfi[
\sqrt{q\beta} (Q + 2 \lambda)]+\piup \erfi[\sqrt{q\beta} (Q + 2 \lambda)]^2-\sqrt{\piup} \erfi[Q \sqrt{q\beta}]\nonumber
\\
{}&\times\Big(\sqrt{q\beta} \big[\re^{q Q^2 \beta} (Q - 2 q Q^3 \beta) + 
\re^{q \beta (Q + 2 \lambda)^2} (Q + 2 \lambda) (-1 + 
2 q \beta (Q + 2 \lambda)^2)\big]\nonumber\\
{}&+ 
2\sqrt{\piup} \erfi[\sqrt{q\beta} (Q + 2 \lambda)]\Big)\Big\} \big/ \Big\{2 \piup \left(\erfi[Q \sqrt{q\beta}] - 
\erfi[\sqrt{q\beta} (Q + 2 \lambda)]\right)^2\Big\}.
\end{align}

\item \textbf{Free energy}\\
Free energy is calculated by:
\begin{align}
F(\beta){}&=-\frac{1}{\beta}\ln Z(\beta),\\
F(\beta){}&=-\frac{1}{\beta}\ln\left(\frac{\re^{-V\beta}\sqrt{\piup}\left(-\erfi[Q\sqrt{q\beta}]+\erfi[(Q+2\lambda)\sqrt{q\beta}]\right)}{4\sqrt{q\beta}}\right).
\end{align}

\item \textbf{Entropy}\\
Entropy is given by:
\begin{align}
S(\beta)= {}&k_{\text{B}}\ln Z(\beta)-k_{\text{B}}\beta\frac{\partial \ln Z(\beta)}{\partial\beta}=k_{\text{B}}\beta^{2}\frac{\partial F(\beta)}{\partial\beta},\nonumber\\
S(\beta)= {}&k_{\text{B}}\Big\{-2\sqrt{q\beta}\Big[\re^{Q^{2}q\beta}Q-\re^{q\beta(Q+2\lambda)^{2}}(Q+2\lambda)\Big]\nonumber\\
{}&+\sqrt{\piup}\Big[1+2V \beta+2\ln Z(\beta)\Big]\left(\erfi[Q\sqrt{q\beta}]-\erfi[(Q+2\lambda)\sqrt{q\beta}]\right)\Big\}.
\end{align}
\end{enumerate}

\newpage
\section{Numerical results and discussions}
\label{sec:6}

\begin{figure}[h]
	\centerline{\includegraphics[width=0.50\textwidth]{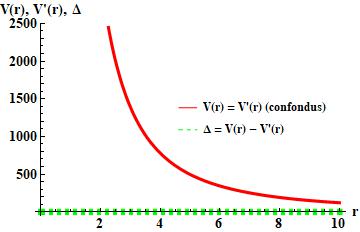}}
	\caption{(Colour online) The variation of potentials as a function of $\beta$ for $V_{0}=V_{5}=4$, $V_{1}=V_{2}=2$ and $V_{3}=-V_{4}=4$.}
	\label{fig-smp1}
\end{figure}

Figure~\ref{fig-smp1} represents the plot of potentials as a function of $r$. The variation of the approximate potential $V'$ closely follows the variation of the actual potential $V$. Both are decreasing exponential functions, which coincide. The fact that these two curves coincide indicates a good approximation. The curve shows the difference between $V$ and $V'$, and it is very close to 0. It practically follows the $x$-axis, confirming the accuracy of the approximation.

\begin{figure}[h]
	\centerline{\includegraphics[width=0.45\textwidth]{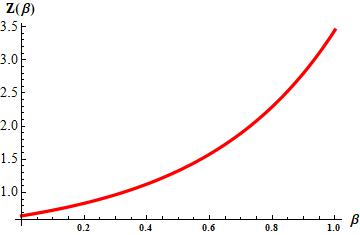}}
	\caption{(Colour online) The variation of $Z(\beta)$ of a few molecules for $Q=V=\lambda=0.6$ and $q=1.8$.}
	\label{fig-smp2}
\end{figure}

\begin{figure}[h]
	\centering
	\begin{subfigure}{0.4\textwidth}
		\includegraphics[width=\textwidth]{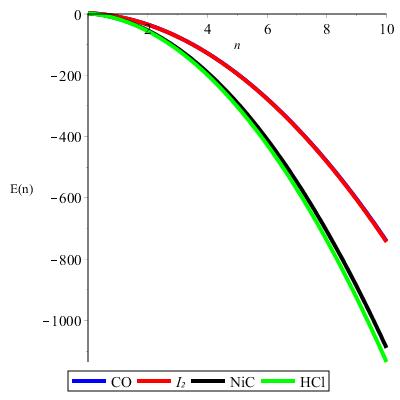}
		\caption{}
		\label{fig:sub1}
	\end{subfigure}
	\hspace{0.5cm}
	\begin{subfigure}{0.4\textwidth}
		\includegraphics[width=\textwidth]{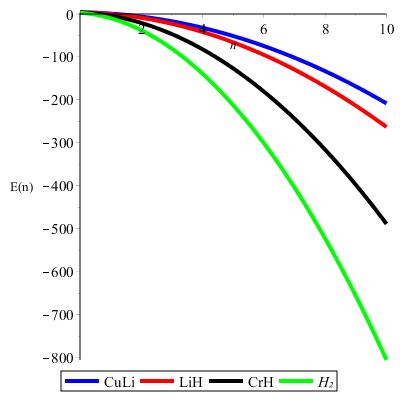}
		\caption{}
		\label{fig:sub2}
	\end{subfigure}
	
	\caption{(Colour online) The variation of $E(n)$ of a few molecules for $V_{0}=V_{5}=4$, $V_{1}=V_{2}=2$, $V_{3}=-V_{4}=4$ and $\ell=0$.}
	\label{fig-smp3}
\end{figure}

Figure~\ref{fig-smp2} shows the variation of the partition function $Z$ as a function of $\beta$. It starts near the origin (0,0) and exhibits exponential (non-linear) growth with increasing $\beta$.
Figure~\ref{fig-smp3} depicts the variations in the energy as a function of the quantum number $n$ for certain molecules. All curves show a decreasing trend: the energy decreases as the quantum number $n$ increases. This suggests that higher energy states are less stable. The trend aligns with the quantum model, where higher energy levels correspond to orbits further away from the nucleus.

\begin{figure}[H]
	\centering
	\begin{subfigure}{0.4\textwidth}
		\includegraphics[width=\textwidth]{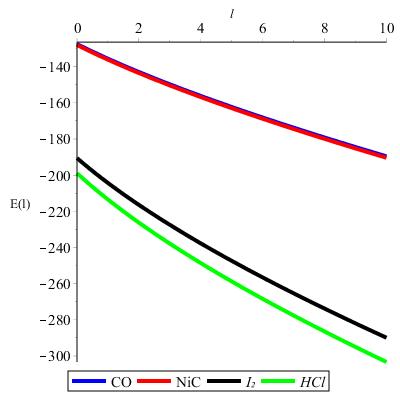}
		\caption{}
		\label{fig:sub3}
	\end{subfigure}
	\hspace{0.5cm}
	\begin{subfigure}{0.4\textwidth}
		\includegraphics[width=\textwidth]{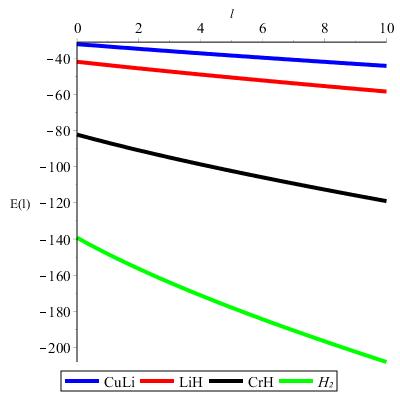}
		\caption{}
		\label{fig:sub4}
	\end{subfigure}
	
	\caption{(Colour online) The variation of $E(\ell)$ of a few molecules for $V_{0}=V_{5}=4$, $V_{1}=V_{2}=2$, $V_{3}=-V_{4}=4$ and $n=4$.}
	\label{fig-smp4}
\end{figure}

\begin{figure}[H]
	\centering
	\begin{subfigure}{0.4\textwidth}
		\includegraphics[width=\textwidth]{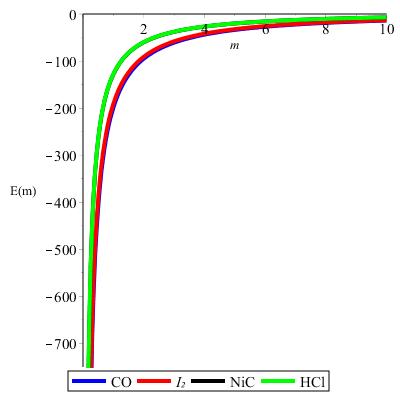}
		\caption{}
		\label{fig:sub5}
	\end{subfigure}
	\hspace{0.5cm}
	\begin{subfigure}{0.4\textwidth}
		\includegraphics[width=\textwidth]{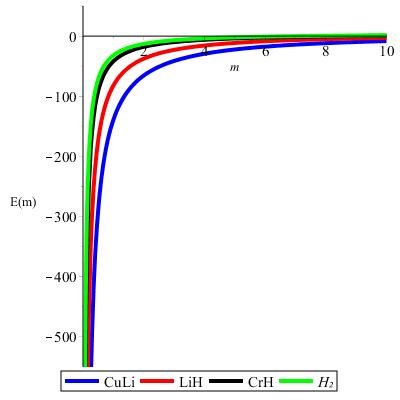}
		\caption{}
		\label{fig:sub6}
	\end{subfigure}
	
	\caption{(Colour online) The variation of $E(m)$ of a few molecules for $V_{0}=V_{5}=4$, $V_{1}=V_{2}=2$, $V_{3}=-V_{4}=4$, $\ell=0$ and $n=4$.}
	\label{fig-smp5}
\end{figure}

\begin{figure}[H]
	\centering
	\begin{subfigure}{0.4\textwidth}
		\includegraphics[width=\textwidth]{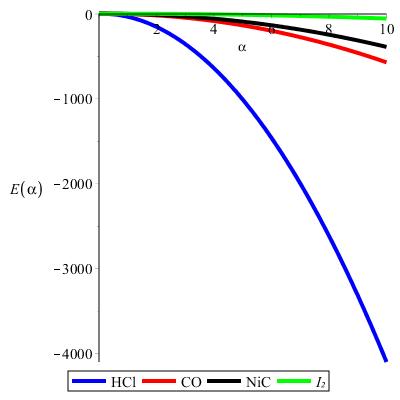}
		\caption{}
		\label{fig:sub7}
	\end{subfigure}
	\hspace{0.5cm}
	\begin{subfigure}{0.4\textwidth}
		\includegraphics[width=\textwidth]{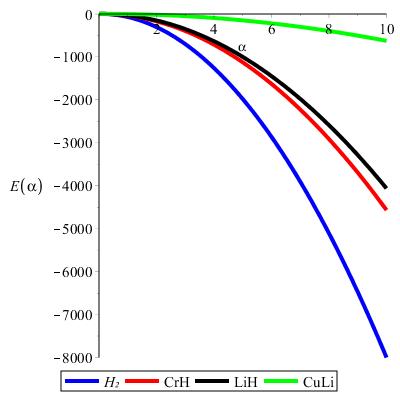}
		\caption{}
		\label{fig:sub8}
	\end{subfigure}
	
	\caption{(Colour online) The variation of $E(\alpha)$ of a few molecules for $V_{0}=V_{5}=4$, $V_{1}=V_{2}=2$, $V_{3}=-V_{4}=4$, $\ell=0$ and $n=4$.}
	\label{fig-smp6}
\end{figure}

Figure~\ref{fig-smp4} illustrates the variation of energy $E$ for certain molecules as a function of angular momentum~$\ell$. All curves exhibit a monotoniously decreasing trend in energy as angular momentum $\ell$ increases. The higher is the angular momentum, the lower is the energy of the molecules. This consistency aligns with quantum mechanics, where lower energy levels correspond to faster rotations and more stable states.
Figure~\ref{fig-smp5} demonstrates the variation of energy $E$ as a function of reduced mass $m$. For each molecule, initially, it increases rapidly in a non-linear fashion for small values of $m$, and then slows down as $m$ increases.
Figure~\ref{fig-smp6} shows the variations in the energy of certain molecules as a function of the inverse of the potential range $\alpha$. In all cases, the energy decreases as the inverse of the potential range increases.

\begin{figure}[H]
	\centering
	\begin{subfigure}{0.4\textwidth}
		\includegraphics[width=\textwidth]{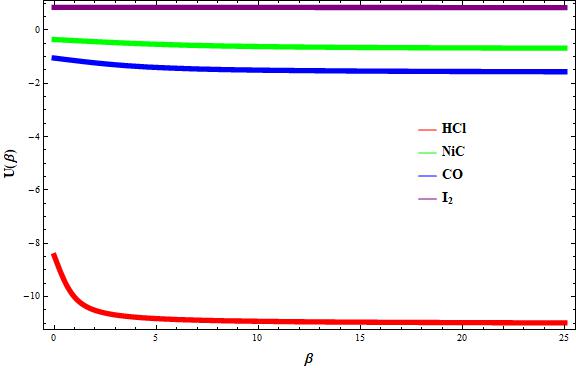}
		\caption{}
		\label{fig:sub9}
	\end{subfigure}
	\hspace{0.5cm}
	\begin{subfigure}{0.4\textwidth}
		\includegraphics[width=\textwidth]{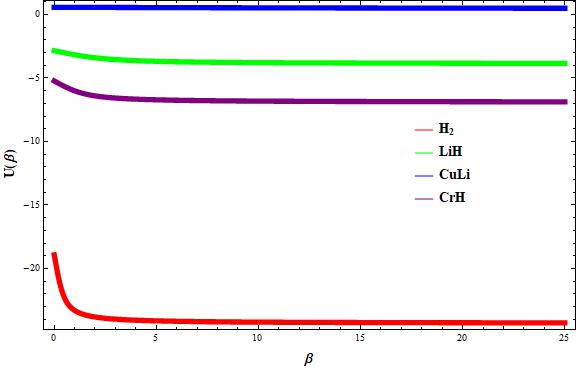}
		\caption{}
		\label{fig:sub10}
	\end{subfigure}
	
	\caption{(Colour online) The variation of $U(\beta)$ of a few molecules for $Q=2$, $V=1$ and $\lambda=0.3$.}
	\label{fig-smp7}
\end{figure}

Figure~\ref{fig-smp7} shows the variations in the average energy of certain molecules as a function of $\beta$. The stability of a molecule is related to its average energy: the lower is the energy, the more stable is the molecule. For the potential $V_{\text{GHHWSP}}(r)$, we deduce that HCl is the most stable, followed by CO, NiC, and I$_{2}$, which respectively have lower average energies in this order. Also H$_{2}$ is more stable, followed respectively by CrH, LiH, and CuLi, which also have lower average energies in this order.

\begin{figure}[h]
	\centering
	\begin{subfigure}{0.4\textwidth}
		\includegraphics[width=\textwidth]{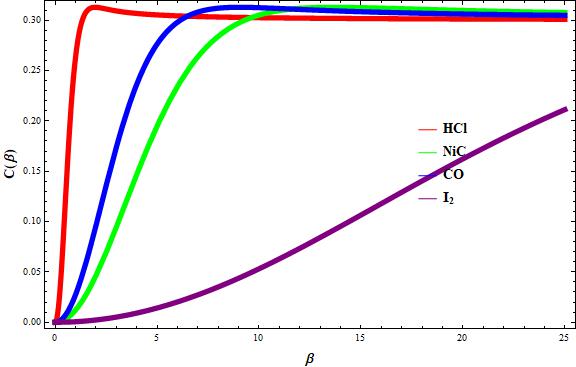}
		\caption{}
		\label{fig:sub11}
	\end{subfigure}
	\hspace{0.5cm}
	\begin{subfigure}{0.4\textwidth}
		\includegraphics[width=\textwidth]{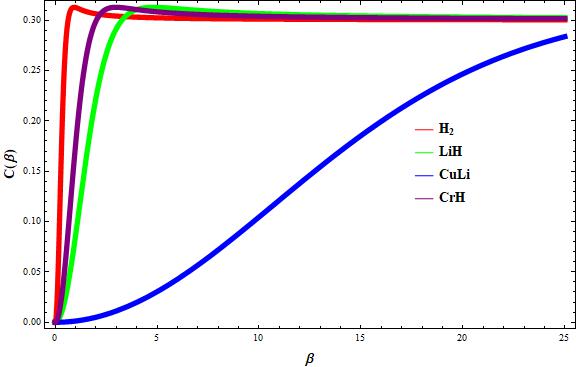}
		\caption{}
		\label{fig:sub12}
	\end{subfigure}
	
	\caption{(Colour online) The variation of $C(\beta)$ of a few molecules for $Q=2$, $V=1$ and  $k=\lambda=0.3$.}
	\label{fig-smp8}
\end{figure}

\begin{figure}[h]
	\centering
	\begin{subfigure}{0.4\textwidth}
		\includegraphics[width=\textwidth]{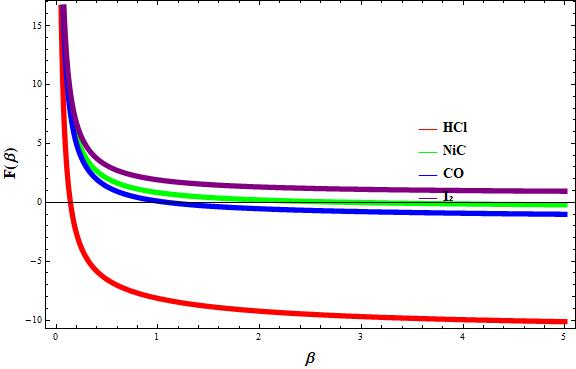}
		\caption{}
		\label{fig:sub13}
	\end{subfigure}
	\hspace{0.5cm}
	\begin{subfigure}{0.4\textwidth}
		\includegraphics[width=\textwidth]{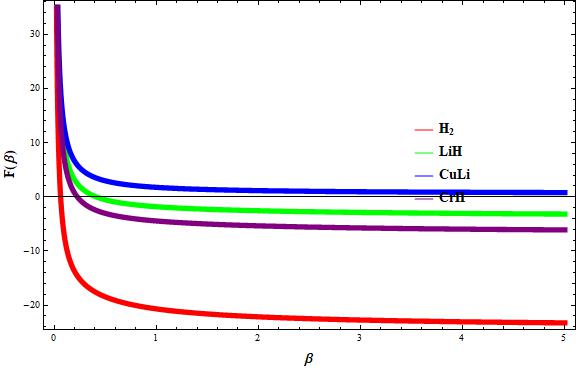}
		\caption{}
		\label{fig:sub14}
	\end{subfigure}
	
	\caption{(Colour online) The variation of $F(\beta)$ of a few molecules for $Q=2$, $V=1$ and $\lambda=0.3$.}
	\label{fig-smp9}
\end{figure}

Figure~\ref{fig-smp8} shows the variation of the specific heat capacity as a function of $\beta$. The molecules HCl, NiC, CO, H$_{2}$, LiH, and CrH exhibit similar behaviors: initially, their specific heat capacity increases quickly, and then it stabilizes. By contrast, the specific heat capacity of I$_{2}$ and CuLi gradually increases  with the increase of $\beta$.
Figure~\ref{fig-smp9} represents the traces of the free energy as a function of $\beta$ for certain molecules. For each molecule, the free energy decreases with the increase of $\beta$. Specifically, for HCl and H$_{2}$ molecules, it decreases more rapidly, quickly reaching stabilization, compared to the other molecules.

\begin{figure}[h]
	\centering
	\begin{subfigure}{0.4\textwidth}
		\includegraphics[width=\textwidth]{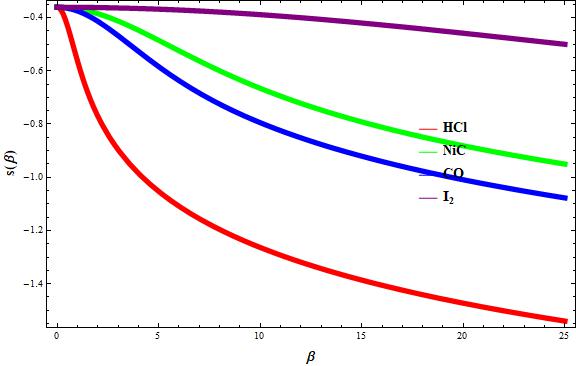}
		\caption{}
		\label{fig:sub15}
	\end{subfigure}
	\hspace{0.5cm}
	\begin{subfigure}{0.4\textwidth}
		\includegraphics[width=\textwidth]{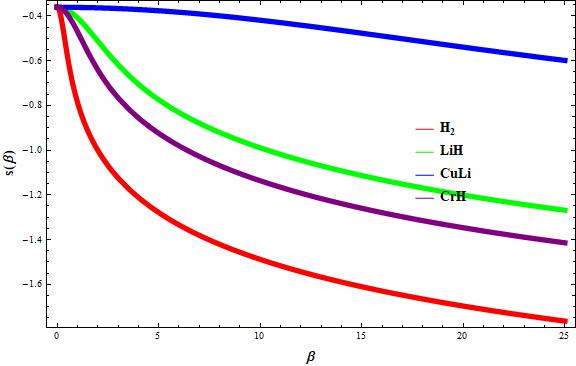}
		\caption{}
		\label{fig:sub16}
	\end{subfigure}
	
	\caption{(Colour online) The variation of $S(\beta)$ of a few molecules for $Q=2$, $V=1$, $k=0.3$ and $\lambda=0.3$.}
	\label{fig-smp10}
\end{figure}

Figure~\ref{fig-smp10} illustrates the variation of the entropy of certain molecules as a function of $\beta$. In all cases, the entropy decreases with increasing $\beta$. Notably, entropy decreases more rapidly with increasing $\beta$ in HCl and H$_{2}$ molecules than in the other molecules. This reduction in disorder occurs as $\beta$ increases, which corresponds to a decrease in temperature $T$. The effect is more pronounced for HCl and H$_{2}$ than for the other molecules.

\section{Conclusion}
\label{sec:concl}
In this work, we used the parametric Nikiforov--Uvarov method and the Greene-Aldrich approximation scheme to find the solutions of
the Schr\"{o}dinger equation of the potential $V_{\text{GHHWSP}}(r)$. The corresponding
eigenvalues and eigenfunctions as a function of $n$, $\ell$, $m$ and $\alpha$ were found. We presented some special cases such as generalized hyperbolic potential, the Rosen--Morse  hyperbolic potential, the Eckart  hyperbolic potential. This method allows us to precisely evaluate the eigenvalues and eigenfunctions of certain potential models.
As we know, there is no analytical solution for $\ell$ non-zero, but thanks to the approach used we were able to find the energy as a function of $n$, $\ell$. The model parameters of certain molecules being known (see table~\ref{Tab-smp1}), we chose to calculate the eigenenergies of the molecules HCl, NiC, CO, I$_{2}$, H$_{2}$, LiH, CuLi, and CrH. The motivation for the choice of these molecules is due to the fact that the parameters of these molecules are known. Moreover, they are used in several fields of physics and chemistry  \cite{D79}. Thermodynamic properties are also studied. Our results are consistent with those obtained in
the literature for particular cases. The thermodynamic properties also align with those in the 
literature but exhibit unique behaviors.

\appendix
\section{Appendix}
\label{app}

\renewcommand*\thetable{A.\arabic{table}}

The necessary information about the molecules selected in this work is provided in the table~\ref{Tab-smp1}. 
\begin{table}[H]
	\centering
	\caption{Model parameters for some selected molecules in this study \cite{D71}.}
	\label{Tab-smp1}
	\begin{tabular}{|c|c|c|}
		\hline
		Molecules&$\alpha$ (\AA${}^{-1}$) & $m$ (amu)\\
		\hline
		HCl&1.8677& 0.9801045\\
		\hline
		NiC&2.25297& 9.974265\\
		\hline
		CO&2.2994& 6.8606719\\
		\hline
		I$_{2}$&1.8643& 63.45223502\\
		\hline
		H$_{2}$&1.9426& 0.50391\\
		\hline
		LiH&1.1280& 0.8801221\\
		\hline
		CuLi&1.00818& 6.259494\\
		\hline
		CrH&1.52179& 0.988976\\
		\hline
	\end{tabular}
\end{table}

Table~\ref{Tab-smp2} and table~\ref{Tab-smp3} illustrate the calculated numerical values for certain molecules, ranging from $n = 0$ to 5 and $0 \leqslant \ell \leqslant n$. The numerical value of HCl increases from the ground state ($n = \ell = 0$) up to the state ($n = 2$, $\ell = 0$), and then decreases to the state ($n = \ell = 5$). Similarly, for the CO molecule, it increases from the ground state ($n =\ell = 0$) up to the state ($n = 5$, $\ell = 0$), and then decreases to the state ($n = \ell = 5$). Likewise, for the LiH molecule, it increases from the ground state ($n = \ell = 0$) up to the state ($n = \ell = 3$), and then decreases to the state ($n = \ell = 5$). The CrH molecule increases from the ground state ($n = \ell = 0$) up to the state ($n = \ell = 2$), and then decreases to the state ($n = \ell = 5$). As for the NiC and H$_{2}$ molecules, they increase from the ground state ($n = \ell = 0$) up to the state ($n = \ell = 1$), and then decrease to the state ($n = \ell = 5$). The numerical energy value of the I$_{2}$ and CuLi molecules increases from the ground state. In summary, each molecule possesses its own specific numerical values, all of which are positive here, varying with $n$ for $\ell$ being a constant and simultaneously with $n$ and $\ell$. The H$_{2}$ molecule exhibits a strong fundamental energy compared to the other selected molecules. As for table~\ref{Tab-smp4}, it presents the numerical values of $E$ for different combinations of $n$ and $\ell$, in the case where $m=\hbar=1$. The numerical values of $E$ decrease here with an increase in $n$ and $\ell$, except in two specific cases: when transitioning from the pair ($n$, $\ell$) $=$ (4, 4) to ($n$, $\ell$) $=$ (5, 0), the value of $E$ increased, additionally, the value of $E$ for the pair ($n$, $\ell$) $=$ (4, 4) repeats at the level of the pair (5, 1) before continuing to decrease. The values of $E$ are positive for $n = 0$ and $n = 1$, and negative for the remaining $n$ values. The values of $E$ depend on $n$ and $\ell$.

\begin{table}[H]
	\centering
	\caption{Energy spectrum $E$~(eV) of HCl, NiC, CO, I$_{2}$ for $n$ and $\ell$ arbitrary, $\hbar c=1973.29$~eV{\AA}, $V_{0}=V_{5}=4$, $V_{1}=V_{2}=2$ and $V_{3}=-V_{4}=4$.}
	\label{Tab-smp2}
	\begin{tabular}{|c|c|c||c|c|c|}
		\hline
		$n$&$\ell$&$E$ (HCL)&$E$ (NiC)&$E$ (CO)&$E$ (I$_{2}$)\\
		\hline
		0&0&3.895432223576353&3.923055258184589 &3.863443911158645&3.840812980991506\\
		\hline
		1&0&3.977238145605624&3.999828313366208&3.916325119675707&3.857431264166523 \\
		\hline
		1&1&3.978230001812676&3.999971795966395 &3.916741268138278 &3.857469925596589 \\
		\hline
		2&0&3.999532196
		503512&3.936906948269785&3.956320149536483&3.873133650799820\\
		\hline
		2&1&3.999377373146526 &3.932931700404279&3.956620676442839&3.873170120921895\\
		\hline
		2&2&3.999002375647724&3.924661141125134&3.957218330408955&3.873243042933650\\
		\hline
		3&0&3.962314376270017&3.734291162895319&3.983429000740975&3.887920140891398\\
		\hline
		3&1&3.961012873348993&3.726197184564121&3.983613906091114&3.887954419705482\\
		\hline
		3&2&3.958349606423559&3.709731759683545&3.983980428553988&3.888022959251768\\
		\hline
		3&3&3.954205380903225 &3.684354056958821&3.984522007499936&3.888125723372700\\
		\hline
		4&0&3.865584684905139&3.391980957242812&3.997651673289180 &3.901790734441255\\
		\hline
		4&1&3.863136502420077&3.379768248445921&3.997720957083104&3.901822821947348\\
		\hline
		4&2&3.858184966068011&3.355107957963914&3.997856348042736&3.901886979028165\\
		\hline
		4&3&3.850620948533141&3.317542279821468&3.998051508218979&3.901983169827125\\
		\hline
		4&4&3.840283735617699&3.266412971037536&3.998296968784796&3.902111340574744\\
		\hline
		5&0&3.709343122408877&2.909976331312260&3.998988167181102&3.914745431449392\\
		\hline
		5&1&3.705748260359777&2.893644892049677&3.998941829418809 &3.914775327647495\\
		\hline
		5&2&3.698508454581080&2.860789735966240&3.998846088875199&3.914835102262843\\
		\hline
		5&3&3.687524645031674&2.811036082406071&3.998694830281737 &3.914924719739831\\
		\hline
		5&4&3.672650943810611&2.743845063924313&3.998478917493729&3.915044126760305\\
		\hline
		5&5&3.653697725636183
		&2.65853874864080&3.998186237139443&3.915193252261870\\
		\hline
		
	\end{tabular}
\end{table}

\begin{table}[H]
	\centering
	\caption{Energy spectrum $E$~(eV) of H$_{2}$, LiH, CuLi, CrH for $n$ and $\ell$ arbitrary, $\hbar c=1973.29$~eV{\AA}, $V_{0}=V_{5}=4$, $V_{1}=V_{2}=2$, and  $V_{3}=-V_{4}=4$.}
	\label{Tab-smp3}
	\begin{tabular}{|c|c|c||c|c|c|}
		\hline
		$n$&$\ell$&$E$ (H$_{2}$)&$E$ (LiH)&$E$ (CuLi)&$E$ (CrH)\\
		\hline
		0&0&3.919095198564432&3.874196804526624&3.846917450921436&3.884616609476618 \\
		\hline
		1&0&3.998829179178609&3.940098404330357&3.874392020796937&3.960095153772992\\
		\hline
		1&1&3.999258479930408&3.940757801555560     &3.874499580816594&3.960964321112187\\
		\hline
		2&0&3.953342628310359&3.981826644594360&3.899151410341168&3.996418858370650\\
		\hline
		2&1&3.950275652482520 &3.982189024639254&3.899247785782805&3.996675882974972\\
		\hline
		2&2&3.943879573821263&3.982902197398643&3.899440378851668&3.997160539593435\\
		\hline
		3&0&3.782635545959681&3.999381525318634 &3.921195619554128&3.993587723269594\\
		\hline
		3&1&3.776072293552204&3.999446888183218&3.921280810417746&3.993232605139042\\
		\hline
		3&2&3.762716116700638&3.999566563857461&3.921451036607602 &3.992494767342508\\
		\hline
		3&3&3.742118169174634&3.999718551873660&3.921705987207206&3.991319408644501 \\
		\hline
		4&0&3.486707932126576&3.992763046503177&3.940524648435818&3.951601748469822\\
		\hline
		4&1&3.476648403139460&3.992531392187452&3.940598654721417&3.950634487604397\\
		\hline
		4&2&3.456332128097585&3.992057570776549&3.940746514032265&3.948674155392865\\
		\hline
		4&3&3.425373687257740&3.991320651508912&3.940967920003620&3.945669506634268\\
		\hline
		4&4&3.383216545608756&3.990289473249351&3.941262413477888&3.941544600249777\\
		\hline
		5&0&3.065559786811044&3.961971208147990&3.957138496986238&3.870460933971335\\
		\hline
		5&1&3.052003981244289 &3.961442536651956&3.957201318693817&3.868881530371037\\
		\hline
		5&2&3.024727608012104&3.960375218155906&3.957326811125658&3.865698703744508\\
		\hline
		5&3&2.983408673858418 &3.958749391604433 &3.957514672468763&3.860864764925319\\
		\hline
		5&4&2.927582229793150&3.956535488411083&3.957764450387370&3.854309042578806\\
		\hline
		5&5&2.856659962059760&3.953694494847846&3.958075542482408&3.845938885774320\\
		\hline
		
	\end{tabular}
\end{table}

\begin{table}[H]
	\centering
	\caption{Energy spectrum $E$~(eV) for arbitrary $n$ and $\ell$, for the atomic unit ($\hbar =m=1$), $\alpha=1$, $V_{0}=V_{5}=4$, $V_{1}=V_{2}=2$ and $V_{3}=-V_{4}=4$.}
	\label{Tab-smp4}
	\begin{tabular}{|c|c|c|}
		\hline
		$n$&$\ell$&$E$ (eV)\\
		\hline
		0&0&3.9089870641907707\\
		\hline
		1&0&1.0556972620286720\\
		\hline
		1&1&0.28696191462993648\\
		\hline
		2&0&-5.7975925401334258\\
		\hline
		2&1&-7.1632036500993141\\
		\hline
		2&2&-9.8138804890362970\\
		\hline
		3&0&-16.650882342295525\\
		\hline
		3&1&-18.613369214828563\\
		\hline
		3&2&-22.326303802064096\\
		\hline
		3&3&-27.538264535626617\\
		\hline
		4&0&-31.504172144457627\\
		\hline
		4&1&-34.063534779557813\\
		\hline
		4&2&-38.838727115091892\\
		\hline
		4&3&-45.422411232412479\\
		\hline
		4&4&-53.513700344287066\\
		\hline
		5&0&-50.357461946619722\\
		\hline
		5&1&-53.513700344287066\\
		\hline
		5&2&-59.351150428119695\\
		\hline
		5&3&-67.306557929198334\\
		\hline
		5&4&-76.963865909016320\\
		\hline
		5&5&-88.063922166111013\\
		\hline
		
	\end{tabular}
\end{table}


\begin{thebibliography}{99}
	
\bibitem{D1}  Greiner W., Relativistics Quantum Mechanics, Springer Berlin, Heidelberg, 2000.
\bibitem{D7} Schr\"{o}dinger E., 	Ann. Phys., 1926,
{\bf 79}, 361--376, \doi{10.1002/andp.19263840404}.
\bibitem{As7} Sun W., Liu Y., Li M., Cheng Q., Zhao L., Energy, 2023, {\bf 269}, 127001, \doi{10.1016/j.energy.2023.127001}.
\bibitem{As10} Dong S. H., Cruz-Irisson M., J. Math. Chem., 2012, {\bf 50}, 
881, \doi{10.1007/s10910-011-9931-3}.
\bibitem{As12} Khordad R., Sedehi H. R. R., J. Low Temp. Phys., 2018, {\bf 190}, 200, \doi{10.1007/s10909-017-1831-x}.
\bibitem{As13} Inyang E. P., William E. S., Omugbe E., Inyang E. P., Ibanga E. A., Ayedun F.,
Akpan I. O., Ntibi J. E.,  	Rev.~Mex.~Fis., 2022, {\bf 68}, 020401, \doi{10.31349/revmexfis.68.020401}. 
\bibitem{As16} Njoku I. J., Onyenegecha C. P., Okereke C. J., Opara A. I., Ukewuihe U. M., 
Nwaneho F. U., Results Phys., 2021, {\bf 24}, 104208, \doi{10.1016/j.rinp.2021.104208}.
\bibitem{D51}  Demirci M., Sever R., Eur. Phys. J. 
Plus, 2023, {\bf 138}, 409, \doi{10.1140/epjp/s13360-023-04030-0}.

\bibitem{As19}   Ramantswana M., Rampho G. J., Edet C. O., Ikot A. N., Okorie U. S., Qadir K. W., Abdullah H. Y., Phys. Open, 2023, {\bf 14}, 100135, \doi{10.1016/j.physo.2022.100135}. 
\bibitem{As20} Oluwadare O. J., Oyewumi K. J., Abiola T. O., Indian J. Phys., 2022, {\bf 96}, 1921, \doi{10.1007/s12648-021-02139-5}.

\bibitem{As21} Wang  C. W., Wang J., Liu Y. S., Li J., Peng X. L.,  Jia C. S., Zhang L. H., Yi L. Z., Liu J. Y., Li C. J., Jia X., J.~Mol.~Liq., 2021, 
{\bf 321}, 114912, \doi{10.1016/j.molliq.2020.114912}.
\bibitem{A27} Okon I. B., Omugbe E., Antia A. D., Onate C. A., Akpabio L. E.,  Osafile O., Sci. Rep., 2021, {\bf 11}, 892,\\ \doi{10.1038/s41598-020-77756-x}.

\bibitem{A28} Edet  C. O., Okorie U. S., Osobonge G., Ikot A. N., Rampho G. J,  Sever R., J. Math. Chem., 2020, {\bf 58}, 989--1013, \\
\doi{10.1007/s10910-020-01107-4}.

\bibitem{A29} Ikot A. N., Chukwuocha E. O., Onyeaju M. C., Onate C. A., Ita B. I., Udoh M. E., Pramana, 2018, {\bf 90}, 22, \\ \doi{10.1007/s12043-017-1510-0}.

\bibitem{A31} Okorie  U. S., Ibekwe E. E., Ikot A. N., Onyeaju M. C., Chukwuocha E. O., J. Korean Phys. Soc., 2018, {\bf 73}, 1211--1218, \doi{10.3938/jkps.73.1211}.
\bibitem{A32} Okorie U. S., Ikot A. N., Onyeaju M. C., Chukwuocha E. O., J. Mol. Model., 2018, {\bf 24}, 1--12, \doi{10.1007/s00894-018-3811-8}.
\bibitem{A23} Okon I. B., Popoola O. O., Omugbe E., Antia A. D., Isonguyo C. N., Ituen E. E., Comput. Theor. Chem., 2021, {\bf 1196}, 113132,
\doi{10.1016/j.comptc.2020.113132}.

\bibitem{A34}  Omugbe E., Osafile O. E., Okon I. B., Eur. Phys. J. Plus, 2021, {\bf 136}, 740, \doi{10.1140/epjp/s13360-021-01712-5}.
\bibitem{D81} Isonguyo C. N., Okon I. B., Antia A. D., Oyewumi K. J., Omugbe E., Onate C. A., Joshua R. U., Udoh M. E., Ituen E. E., Aruajo J. P., Front. Phys., 2022, {\bf 10}, 962717, \doi{10.3389/fphy.2022.962717}.

\bibitem{A} Okon I. B., Isonguyo C. N., Onate C. A., Antia A. D., Purohit K. R., Ekott E. E., Essien K. E., William E. S., Asuquo N. E., Preprint \arxiv{10.48550/arXiv.2304.08219}, 2023.

\bibitem{As} Emeje K. O., Onate C. A., Okon I. B., Omugbe E., Eyube E. S., Olanrewaju D. B.,  Aghemenloh E., 	J. Low Temp. Phys., 2024, {\bf 215}, 109, \doi{10.1007/s10909-024-03074-5}.


\bibitem{E77} Ikhdair S. M., Sever R., J. Mol. Struct. THEOCHEM, 2007, {\bf 809}, 103, \doi{10.1016/j.theochem.2007.01.019}.
\bibitem{E80} Edet C. O., Amadi P. O., Okorie U. S., Tas A., Ikot A. N., Rampho G., Rev. Mex. Fis., 2020, {\bf 66}, 824, \\ \doi{10.31349/RevMexFis.66.824}.

\bibitem{D53} Qiang W. C., Li K., Chen W. L., J. Phys. A: Math. Theor., 2009, {\bf 42}, 205306, \doi{10.1088/1751-8113/42/20/205306}.

\bibitem{D55} E\u{g}rifes H., Demirhan D., B\"{u}y\"{u}kkili\c{c} F., Phys. Scr., 1999, {\bf 60}, 195, \doi{10.1238/Physica.Regular.060a00195}.
\bibitem{D41}  Okorie U. S., Ikot A. N., Edet C. O., Rampho G. J., Sever R., Akpan I. O., J. Phys. Commun., 2019, {\bf 3}, 095015,\\ \doi{10.1088/2399-6528/ab42c6}.


\bibitem{D69} Ahmadov A. I., Demirci M., Mustamin M. F., Aslanova S. M., Orujova M. Sh., Eur. Phys. J. Plus., 2021, {\bf 136}, 208, \doi{10.1140/epjp/s13360-021-01163-y}.
\bibitem{D115}  Abramovitz~M., Stegun~I.~A. (Eds.), Handbook of Mathematical Functions with
Formulas, Graphs, and Mathematical Tables, National Bureau of Standards
Applied Mathematics Series, Vol.~55, U.S. Government Printing Office,
Washington, D.C., 1964.
\bibitem{D79} Oyewumi K. J., Sen K. D., J. Math. Chem., 2012, {\bf 50}, 1039--1059, \doi{10.1007/s10910-011-9967-4}.
	
\bibitem{D71} Oyewumi K. J., Oluwadare O. J., Sen K. D., Babalola O. A., J. Math. Chem., 2013, {\bf 51}, 976, \doi{10.1007/s10910-012-0123-6}.	
	

\end{thebibliography}

\ukrainianpart

\title{Власні стани та термодинамічні властивості узагальненого гіперболічного потенціалу Хюльтена та Вудса--Саксона}
\author{Й.~М. Ассімю\refaddr{label1}, 
	С.~Т. Даніел\refaddr{label1, label2},
	Г.~Іссуфу\refaddr{label1,label3},
	Д.~Ф. Ансельме\refaddr{label4},
	Г.~Й.~Х.~Авоссеву\refaddr{label1}
}
\addresses{
	\addr{label1} Інститут фізико-математичних наук, Дангбо, Бенін
	\addr{label2} Політехнічна школа Абомей-Калаві, Бенін
	\addr{label3} Факультет прикладних наук університету Доссо, Нігер
	\addr{label4} Національний університет інженерії, технологій та математики, Абомей, Бенін
}

\makeukrtitle

\begin{abstract}
	\tolerance=3000%
	Представлено розв’язки рівняння Шредінгера та термодинамічні властивості узагальненого гіперболічного потенціалу Хюлтена та Вудса-Саксона. Власні значення та власні функції знайдено за допомогою параметричного методу Нікіфорова--Уварова. Затравочні енергії молекул HCl, NiC, CO, I$_2$, H$_2$, LiH, CuLi і CrH розраховано для певних значень $n$ та $\ell$. Вони є додатними та близькими до енергії основного стану ($n= \ell= 0$), але стають від’ємними при $n=2$. Власні значення енергії зменшуються зі збільшенням $n$, $\ell$, $\alpha$, але зростають зі збільшенням $m$, що підтверджує відомі з наукової літератури результати. Отримана енергія використовувалась для розрахунку статистичної суми, за допомогою якої можна обчислити такі термодинамічні властивості, як внутрішня та вільна енергії, питома теплоємність та ентропія. Числові результати отримано для узагальненого гіперболічного потенціалу Хюлтена та Вудса-Саксона. Показано, що безлад у системі зменшується з пониженням температури, при чому це зменшення відбувається швидше для молекул HCl і H$_{2}$.
	\keywords власні стани, термодинамічні властивості, параметричний метод Нікіфорова--Уварова, гіперболічні потенціали Хюлтена та Вудса-Саксона
	
\end{abstract}

\lastpage
\end{document}